\pdfoutput=1
\documentclass[11pt]{article}

\usepackage[]{ACL2023}

\usepackage{times}
\usepackage{latexsym}

\usepackage{bm}
\usepackage{times}
\usepackage{amsmath}
\usepackage{amsfonts}
\usepackage{amssymb}
\usepackage{mathtools}
\usepackage{ctable}
\usepackage{multirow}
\usepackage{stmaryrd}
\usepackage{tcolorbox}
\usepackage{makecell}
\usepackage{cite}
\usepackage{algorithmic}
\usepackage{subcaption}
\usepackage{graphicx}
\usepackage{textcomp}
\usepackage{xcolor}
\usepackage{hyperref}
\usepackage{url}
\newcommand\blfootnote[1]{%
  \begingroup
  \renewcommand\thefootnote{}\footnote{#1}%
  \addtocounter{footnote}{-1}%
  \endgroup
}
\hypersetup{
    colorlinks,
    linkcolor={red!50!black},
    citecolor={blue!50!black},
    urlcolor={blue!85!black}
}

\usepackage[T1]{fontenc}

\usepackage[utf8]{inputenc}

\usepackage[capitalize,nameinlink,noabbrev]{cleveref}

\usepackage{microtype}
\usepackage{inconsolata}
\newcommand{\name}{LMRec}

\title{Pivotal Role of Language Modeling in Recommender Systems: \\ Enriching Task-specific and Task-agnostic Representation Learning}
\author{Kyuyong Shin$^{\dagger\ddagger\S}$~~~Hanock Kwak$^{\dagger\S}$~~~Wonjae Kim$^{\ddagger}$~~~Jisu Jeong$^{\dagger\ddagger}$ \vspace{0.5mm}\\\textbf{Seungjae Jung}$^{\dagger}$~~~\textbf{Kyung-Min Kim}$^{\dagger\ddagger}$~~~\textbf{Jung-Woo Ha}$^{\ddagger}$~~~\textbf{Sang-Woo Lee}$^{\ddagger}$  \\\\
NAVER$^{\dagger}$~~~NAVER AI Lab$^{\ddagger}$
}

\begin{document}
\maketitle
\begin{abstract}
Recent studies have proposed unified user modeling frameworks that leverage user behavior data from various applications.
Many of them benefit from utilizing users' behavior sequences as plain texts, representing rich information in any domain or system without losing generality. 
Hence, a question arises: Can \textit{language modeling} for user history corpus help improve recommender systems? 
While its versatile usability has been widely investigated in many domains, its applications to recommender systems still remain underexplored.
We show that language modeling applied directly to \textit{task-specific user histories} achieves excellent results on diverse recommendation tasks.
Also, leveraging additional \textit{task-agnostic user histories} delivers significant performance benefits.
We further demonstrate that our approach can provide promising transfer learning capabilities for a broad spectrum of real-world recommender systems, even on unseen domains and services. 
\blfootnote{$^{\S}$Both authors contributed equally to this research. Correspondence to: $\text{<ky.shin@navercorp.com>}$.}
\end{abstract}

\section{Introduction} \label{sec:intro}
\begin{figure*}[t]
\begin{center}
\centerline{\includegraphics[width=0.90\textwidth]{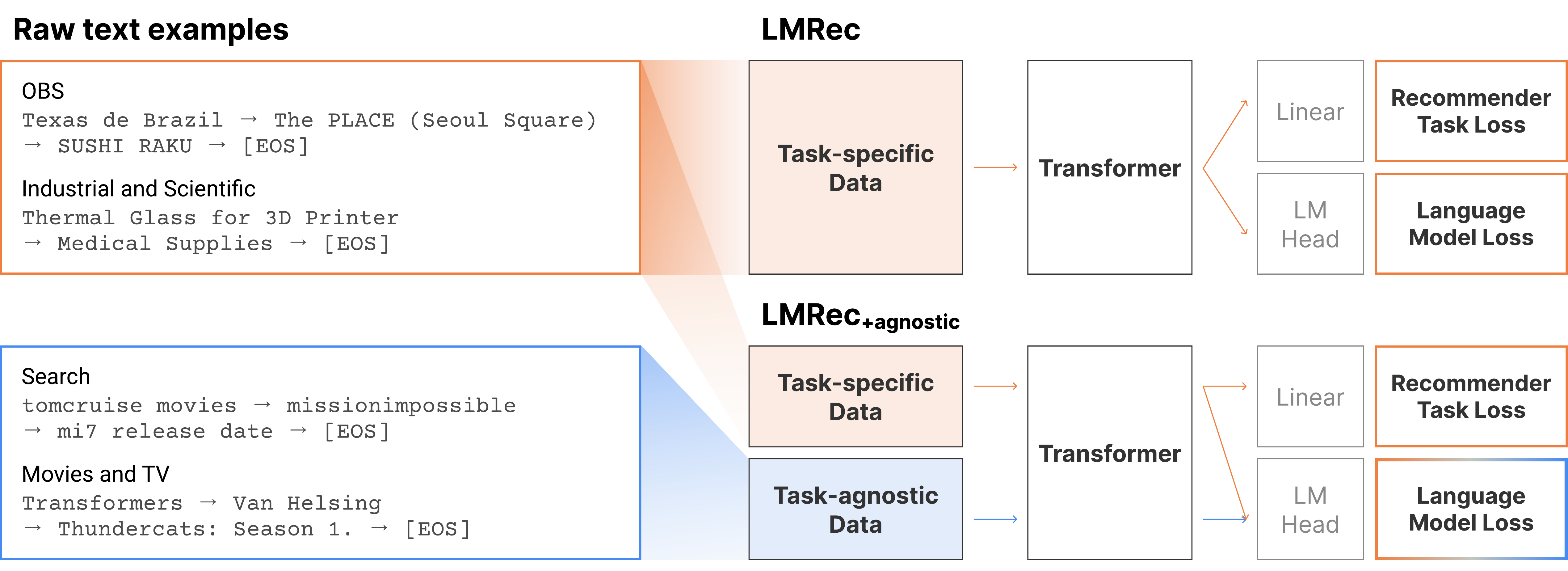}}
\caption{Schematic overview of \name. Task-specific data refers to the user history data of the target recommendation task. Task-agnostic data is collected from other services that do not overlap with target tasks. \textbf{(Left)} We append $\texttt{[EOS]}$ token at the end of every input and use the last layer hidden vector of $\texttt{[EOS]}$ token as a user feature. \textbf{(Right)} The transformer layers are shared across language modeling and recommendation tasks, while the top linear layers are not. \name$_{\tiny\text{+agnostic}}$ incorporates additional task-agnostic data, which delivers large performance benefits.}
\label{fig:overview}
\end{center}
\end{figure*} 
Recent advances in user modeling have focused on constructing unified user models to be directly adapted to diverse applications. Many of them leverage natural language or plain text data, which enables general-purpose applicability among various domains and systems~\citep{qiu2021u, gu2021exploiting, geng2022recommendation, cui2022m6, hou2022towards, shin2021scaling}.
These strategies pave a much more efficient way for service owners to quickly adapt to various task scenarios by tuning one single model, bringing performance improvement across whole systems in parallel.

Based on the recent explosions of sequence prediction models in many domains~\citep{chen2020generative, NEURIPS2020_1457c0d6, ramesh2021zero, chen2021evaluating, borsos2022audiolm}, it is natural to ask whether recommender systems can benefit from representation trained by token sequence prediction, i.e., \textit{language modeling}.  
Moreover, several works have provided deep insights into why and how language models help address downstream classification tasks~\citep{gururangan-etal-2020-dont, saunshi2021a, wei2021why, karouzos-etal-2021-udalm, krishna2022downstream}. 

Some recent studies confirm that continued pretraining of language model on few task-specific data drawn from \textit{the target task distribution}, or data similar to a target domain can provide significant benefits to solve downstream classification tasks~\citep{gururangan-etal-2020-dont, lee2020biobert, karouzos-etal-2021-udalm}. Interestingly,~\citet{krishna2022downstream} go further and validate that language models \textit{trained from scratch} on task-specific or task-agnostic data\footnote{Other studies, such as \citet{gururangan-etal-2020-dont} and \citet{krishna2022downstream}, use the term “domain-specific data” or “cross-data” to represent task-irrelevant corpus that is not webtext data. However, we use the term “task-agnostic data” to generally refer to data from other downstream tasks.}---data from \textit{other downstream tasks}---can rival standard webtext language models.
Another line of research provides mathematical explanations of how language model pretraining can improve performances on downstream tasks~\citep{saunshi2021a, wei2021why}. More specifically, \citet{saunshi2021a} reformulate classification tasks as sentence completion tasks, thus demonstrating that linear classification using output features from fixed GPT-2~\citep{radford2019language}, i.e., no finetuning, also guarantees to solve sentence classification tasks.

Motivated by these works, we introduce a new method called \textbf{\name}, which jointly trains \textbf{L}anguage \textbf{M}odel and \textbf{Rec}ommendation task objectives from user behavior histories transformed as plain text format. 
As illustrated in~\cref{fig:overview}, our approach is conceptually simple but practically effective. 
We first investigate if the recommender system jointly trained with the language modeling objective on \textit{task-specific data} can enrich the user/item representations, thus providing better generalization even for unseen downstream tasks (\cref{tab:eval} and \ref{tab:tl-eval}). We then further verify that additional \textit{task-agnostic data} can help across the various recommendation tasks, especially when using the task-agnostic data as a user feature (\cref{fig:corpus_abl}). 
As a result, our methods significantly outperform all the baselines on all tasks, including three public benchmarks and three real-world datasets from different application service domains, and online A/B experiments.
Moreover, the pretrained \name~shows a promising ability to perform downstream transfers flexibly with simple feature-based transfer learning.
We also explore several aspects of how the language modeling regime affects the model quality under various conditions, including transfer learning, corpus ablation, and model sizes.  

Our major findings are as follows:

\noindent\textbf{Jointly training language modeling and recommendation task objectives improve recommender systems.}
Language modeling on the user history can produce rich user/item representations for diverse applications. 
These results are consistent with the effect of task-adaptive pretraining in the previous research~\citep{gururangan-etal-2020-dont, karouzos-etal-2021-udalm, krishna2022downstream}.
Furthermore, our approach also boosts the transfer learning capability of the recommendation model.
Extensive experimental results show the efficacy of our approach compared to training without language model objectives (\cref{tab:eval} and \ref{tab:tl-eval}). 

\noindent\textbf{Language modeling on task-agnostic data provides strong results on user representation learning.}
Consistent with prior work~\citep{gururangan-etal-2020-dont,  krishna2022downstream}, language modeling on additional \textit{task-agnostic data} alleviates overfitting to a specific history corpus and benefits the learning of robust text representations (\cref{tab:eval} and \ref{tab:tl-eval}).
We explore how language model pretraining on the diverse task-agnostic data affects transfer learning performances, by comparing with models pretrained on different domain corpora  (\cref{fig:corpus_abl}). 

\noindent\textbf{Virtues of more user data.}
Recent studies argue that increasing information on user data should be treated as a top priority for improving recommendation performances~\citep{shin2021one4all, ardalani2022understanding}. 
We collect additional user data matched with downstream task users based on user IDs and incorporate them as an additional user feature. \cref{tab:tl-eval} verifies the data scaling strategy has shown to be beneficial to our models.

\section{Approach} \label{sec:approach}
\setlength{\tabcolsep}{5.5pt}
\ctable[
% 	captionskip = 0pt,
    caption = {Linear probe results on downstream recommendation tasks of language model (LM) embeddings pretrained with different source corpora. We pretrain LMs on three datasets: generic webtext corpora (LM$_{\tiny\text{webtext}}$), task-agnostic user history (LM$_{\tiny\text{agnostic}}$), and task-specific user data (LM$_{\tiny\text{specific}}$).},
    botcap,
    label=tab:lmonly,
    pos=t,
%    width=tabularx,
%    figure,
    % star,
%  	doinside=\normalsize
    %   doinside=\small
    %  doinside = \footnotesize
    doinside = \scriptsize
    % doinside = \tiny
]{ccccc}{
}{
\toprule 
\multirow{2}{*}{Method} 
& \multicolumn{2}{c}{OBS} 
& \multicolumn{2}{c}{Scientific}  
\\
\cmidrule(l){2-3}
\cmidrule(l){4-5}
& Recall@10 & NDCG@10  & Recall@10 & NDCG@10  \\
\midrule[0.52pt]
\midrule[0.52pt]
LM$_{\tiny\text{webtext}}$    & 0.3135 & 0.1766   & 0.0335   & 0.0131  \\
LM$_{\tiny\text{agnostic}}$   & 0.3142 & 0.1747   & 0.0327   & 0.0126  \\
LM$_{\tiny\text{specific}}$   & 0.3769 & 0.2136   & 0.0417   & 0.0194  \\
\bottomrule
}
\subsection{Language Models Help with Classification Tasks}
The empirical and theoretical analyses from the prior work imply that the learned features from the language models trained with appropriate behavior corpus could help predict user and item interactions in recommender systems~\citep{gururangan-etal-2020-dont, saunshi2021a, krishna2022downstream}. It is also consistent with the results in \cref{tab:lmonly} that language model pretraining with appropriate corpus---related to the downstream task rather than other corpora such as webtext---leads to performance improvement. It is worth mentioning that linear probe results of LM$_{\tiny\text{agnostic}}$ can achieve that of LM$_{\tiny\text{webtext}}$ performance, although task-agnostic data are in a much smaller-scale than webtext data. 
This result strongly motivates our research.

Given a sequence of text tokens of user history, $u=\{h_{1}, . . . , h_{n}\}$ and item text tokens $i=\{g_{1}, . . . , g_{m}\}$, the language model objective $L_{1}$ is to maximize the following negative log-likelihood:
\begin{align}
L_{1} = - \sum_{j=1}\text{log}P(h_{j}|h_{j-k},\ldots,h_{j-1};\mathcal{M}),
\end{align}
where $k$ is the context size, and the conditional probability $P$ is modeled using language model $\mathcal{M}$. Then for the downstream tasks, user and item representations $\bm{z}_u, \bm{z}_i\in R^d$ are computed as follows: 
\begin{gather}
\bm{z}_u = \mathcal{M}(h_{\tiny\texttt{EOS}}|u) \\
\bm{z}_i = \mathcal{M}(g_{\tiny\texttt{EOS}}|i), 
\end{gather}
where $\texttt{EOS}$ denotes the end of the history token. We use a vector that corresponds to $\texttt{[EOS]}$ token at the last layer as a feature~\citep{neelakantan2022text}.
The downstream recommendation task loss, $L_{2}$, of each user-item pair is defined as:  
\begin{gather} \label{eq:4}
p_{u,i} = \frac{1}{1+\text{exp}(-\big<W_{u}\bm{z}_u, W_{i}\bm{z}_i\big>)}, \\
L_{2} = - y\text{log}p_{u,i} - (1-y)\text{log}(1-p_{u,i}), 
\end{gather}
where $y\in\{0, 1\}$ is the label denoting whether the user interacted with an item or not. We use $\big<\cdot, \cdot\big>$ for the dot product. The weight matrices $W_{u}, W_{i}\in R^{d\times d}$ linearly transform the user and item representations, respectively. 
\setlength{\tabcolsep}{5.5pt}
\ctable[
% 	captionskip = 0pt,
    caption = {The SelfPretrain model is first pretrained with task-specific data and then finetuned to downstream tasks, while \name~is jointly training language modeling and recommendation objectives.},
    botcap,
    label=tab:pt-ft,
    pos=t,
%    width=tabularx,
%    figure,
    % star,
%  	doinside=\normalsize
    %   doinside=\small
    %  doinside = \footnotesize
    doinside = \scriptsize
    % doinside = \tiny
]{ccccc}{
}{
\toprule 
\multirow{2}{*}{Method} 
& \multicolumn{2}{c}{OBS} 
& \multicolumn{2}{c}{Scientific}  
\\
\cmidrule(l){2-3}
\cmidrule(l){4-5}
& Recall@10 & NDCG@10  & Recall@10 & NDCG@10  \\
\midrule[0.52pt]
\midrule[0.52pt]
SelfPretrain & 0.4742 & 0.2796 & 0.1068 & 0.0473 \\
\name        & 0.4867 & 0.2940 & 0.1264 & 0.0695 \\
\bottomrule
}

Several works have highlighted that jointly optimizing language modeling during finetuning benefits avoiding catastrophic forgetting~\citep{chronopoulou-etal-2019-embarrassingly, karouzos-etal-2021-udalm}. Inspired by the merits of this strategy, we adopt a joint optimization:
\begin{align} \label{eq:6}
L = L_{1} + \lambda L_{2},
\end{align}
where $L$ is the final joint training loss. We impose weight $\lambda$ on $L_{2}$ loss to prevent the overfitting of recommendation tasks. As illustrated in \cref{fig:overview}, a model that optimizes \cref{eq:6} is denoted as “\textbf{\name}”. The model trained without the language model objective ($L_{1}$) is “\textbf{\name$_{\tiny\text{-lm}}$}”.
The performance comparison between the pretrain-then-finetune model and our approach are presented in~\cref{tab:pt-ft}.

\subsection{Enriching Task-specific and Task-agnostic Representation}

\noindent\textbf{Leveraging task-agnostic data.} Optimizing performances solely on task-specific data would restrict the potential of a unified framework. Therefore, a recent trend in user modeling research is to leverage large quantities of pretraining (or additional) data that are not directly related to the target task~\citep{hou2022towards, shin2021scaling}.

To this end, we introduce “\textbf{\name$_{\tiny\text{+agnostic}}$}”, which utilizes additional task-agnostic data for language model objectives.
This approach increases the generality by mitigating overfitting to a specific history corpus. Consequently, it boosts the learning of robust text representations, thus making \name$_{\tiny\text{+agnostic}}$ universal across various tasks.
As a result, additional task-agnostic data further boost the performance of our default \name~model, which already produces state-of-the-art results in all tasks and metrics. 
\begin{figure}[t]
\begin{center}
\centerline{\includegraphics[width=1.0\columnwidth]{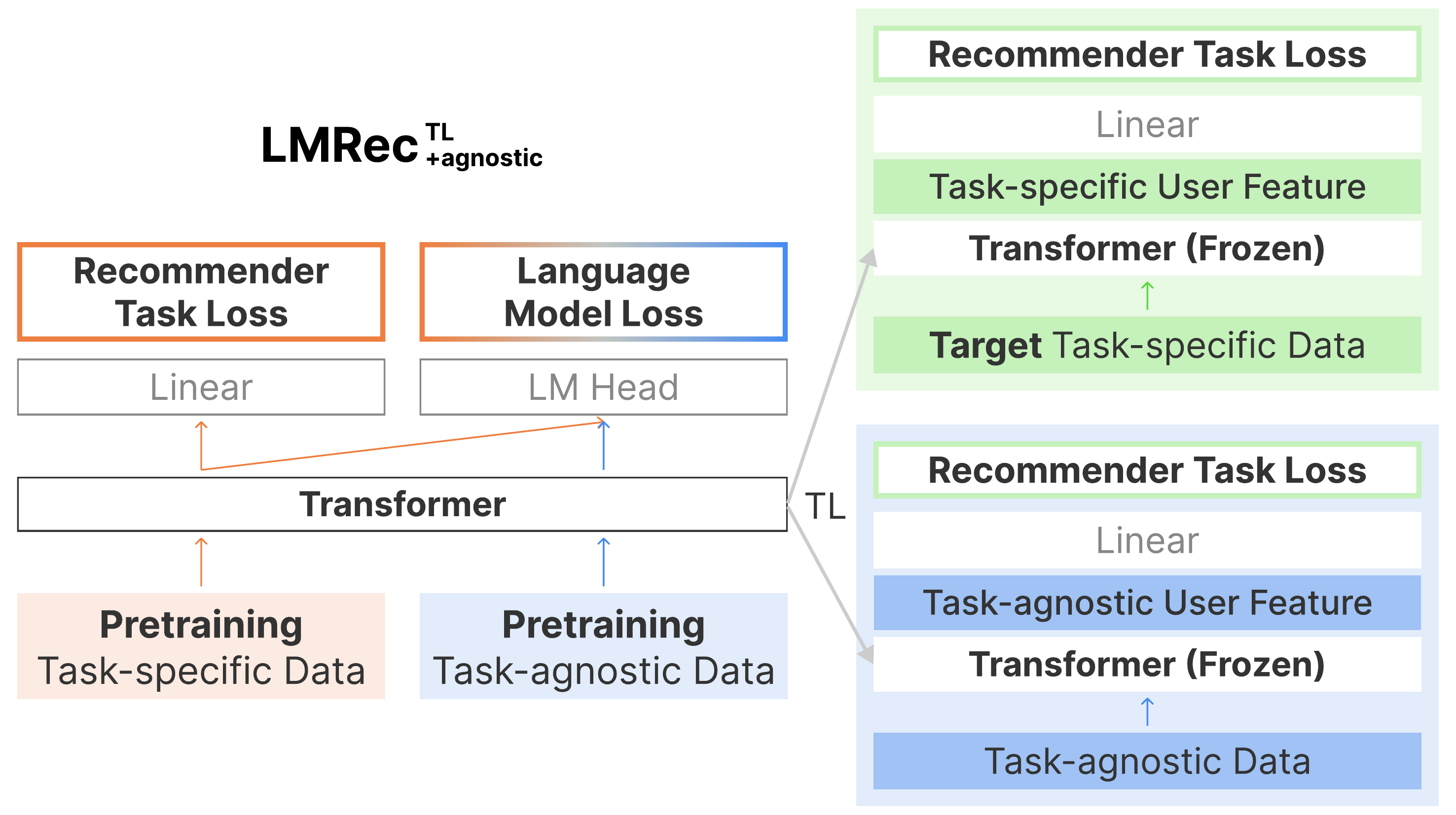}}
\caption{Overall pretraining and feature-based transfer learning procedures of \name$^{\tiny\text{TL}}_{\tiny\text{+agnostic}}$ model. The two types of inputs, i.e., “Target Task-specific Data” and “Task-agnostic Data,” refer to the user history for producing user features. Only the linear layer for the downstream task is trained, while the pretrained transformer parameters are frozen. 
TL denotes transfer learning.}
\label{fig:m2}
\end{center}
\end{figure} 

\noindent\textbf{Transfer learning.} There are several difficulties in applying a unified model to real-world applications: (1) target applications are commonly unknown or undefined during pretraining, (2) user ID cannot be matched across different companies, (3) large-scale recommender systems usually contain millions of users and items, thus it is computationally expensive to finetune the large models to numerous applications directly. To overcome these obstacles, we propose a simple transfer learning framework that can easily and quickly adapt the model to diverse applications.
As visualized in \cref{fig:m2}, we simply plug the target task-specific inputs into the pretrained \name~and compute user/item embeddings to perform a linear probe. We add superscript to the model as “\textbf{\name$^{\tiny\text{TL}}$}” for the transfer learning framework.
The \name$^{\tiny\text{TL}}$ model jointly pretrains multiple tasks, excluding the target downstream task. The final loss to pretrain is as follows:
\begin{align}
L = \sum_{t\in\mathcal{T}_s,\mathcal{T}_a}{L^{t}_{1}}  + \lambda \sum_{t\in\mathcal{T}_s}{L^{t}_{2}},
\end{align}

\noindent where $\mathcal{T}_s$ denotes a set of pretraining recommendation tasks, and $\mathcal{T}_a$ for additional task-agnostic data.
Note that linear layers of pretraining and feature-based transfer learning are separate modules.

\noindent\textbf{Task-agnostic user features.} Leveraging cross-domain data of users for improving recommender systems has been widely discussed~\citep{man2017cross, 10.5555/3367471.3367629, zhu2022personalized, shin2021scaling}. These strategies assume that the underlying user preference in the source and the target domains can be related, and thus learning a common user semantic enhances the recommender system.  
Hence, we utilize additional task-agnostic data, obtained from application services whose user IDs are shared in a company level, as a user feature for target downstream tasks. The difference between task-specific and task-agnostic data in \cref{fig:m2} is only which user features are used for transfer learning. For example, if the target downstream task is ECOMM, models are first pretrained with OBS and OTA, and then use task-specific data of ECOMM to produce task-specific user features. For leveraging the task-agnostic user feature, the pretrained model extracts user features from task-agnostic data, such as Search and News. Components other than user features, such as the pretrained model, downstream architecture (linear layer), and ground truth interacted items of users, are all the same.
We can verify that the transfer learning approach benefits from leveraging additional task-agnostic data as user features, especially when it is recommending for new users (\cref{tab:tl-eval}, \ref{tab:md} and \cref{fig:corpus_abl}). 

\cref{appendix_1} describes the training details of our methods.

\section{Experiments}  \label{sec:exp}
\setlength{\tabcolsep}{4.5pt}
\ctable[
% 	captionskip = 0pt,
    caption = {Statistics of the datasets.},
    botcap,
    label = tab:data,
    pos=t,
%    width=tabularx,
%    figure,
    star,
%  	doinside=\normalsize
      doinside=\small
    %  doinside = \footnotesize
    % doinside = \scriptsize
    % doinside = \tiny
]{ccccccccc}{
}{

\toprule
\multirow{2}{*}{Contents} 
& \multicolumn{4}{c}{In-house} 
& \multicolumn{4}{c}{Public}  
\\
\cmidrule(l){2-5}  
\cmidrule(l){6-9}  
& OBS & OTA & ECOMM & Pretraining & Scientific & Pantry & Online Retail & Pretraining  \\
\midrule
\# of Users         & $300,000$ & $142,051$ & $72,477$  & $10,156,217$  & $8,442$  & $13,101$  & $16,520$  & $1,361,408$ \\
\# of Items         & $42,453$  & $2,485$   & $229,775$ & N/A           & $4,385$  & $4,898$   & $3,469$   & $446,975$   \\
\# of Interact.     & $495,992$ & $177,281$ & $130,859$ & $94,011,305$  & $59,427$ & $126,962$ & $519,906$ & $14,029,229$\\
Avg. history        & $1.5$     & $2.3$     & $5.5$     & $128.7$       & $4.5$    & $8.5$     & $25.6$    & $9.6$\\
Avg. history tokens & $10.3$    & $17.1$    & $116.4$   & $1,222.7$     & $212.5$  & $214.7$   & $206.6$   & $347.3$\\
\bottomrule
}

\subsection{Datasets} \label{sec:exp-data}
To make user behavioral corpora, we consider the behavior description as items, i.e., search queries of search logs, news titles of online news click logs, and content titles of social media click logs. As illustrated in \cref{fig:overview}, we concatenate the behavior logs using the “$\rightarrow$” token. This simple form of a prompt template can have behavior sequences that are very long. Furthermore, separating corpus among multiple services provides flexible transfer learning capabilities by enabling easy proliferation of behaviors and filtering out redundant representation to target applications.
We use Byte-level BPE~\citep{wang2020neural} to tokenize the textual description of each item in the behavior logs.

\noindent\textbf{Task-specific datasets.} We use three in-house datasets in order to assess our approach on various applications and add three public datasets that are predominantly evaluated in recommendation communities. The in-house datasets are built from services of an online booking service (OBS), an online travel agency (OTA), and e-commerce platmform (ECOMM). For public datasets, we select two categories \textit{“Industrial and
Scientific”} (Scientific) and \textit{“Prime Pantry”} (Pantry) from Amazon review datatsets~\citep{ni2019justifying} which are two completely different service domains. We further collect \textit{“Online Retail”}\footnote{https://www.kaggle.com/carrie1/ecommerce-data} dataset from an online retail platform to validate the cross-system transferability of our models.

\noindent\textbf{Task-agnostic datasets.} We construct sufficiently large-scale task-agnostic behavioral corpora for in-house datasets. These datasets are collected over two years and from four behavioral corpora, a search engine (Search), e-commerce (E-comm.), social media platform (SNS), and news website (News). As a result, the in-house dataset contains $10$ million users and $94$ million user history logs, and $12$ billion BBPE tokens. 
Following the experimental setup of UniSRec~\citep{hou2022towards} for public benchmarks, we select the five categories \textit{“Grocery and Gourmet Food”}, \textit{“Home and
Kitchen”}, \textit{“CDs and Vinyl”}, \textit{“Kindle Store”}, and \textit{“Movies and TV”} from Amazon review datasets. These datasets are used as pretraining datasets for pretrain-then-transfer models such as UserBERT~\citep{10.1145/3477495.3531810},  UniSRec~\citep{hou2022towards}, M6-Rec~\citep{cui2022m6}, and CLUE~\citep{shin2021scaling}, while used as additional task-agnostic data for \name$_{\tiny\text{+agnostic}}$ model. 

The details of datasets are outlined in \cref{tab:data}.

\subsection{Experimental Settings}
\noindent\textbf{In-house downstream tasks.} 
The datasets consist of positive pairs $(u, i)$ which means a user $u$ interacted with an item $i$.
The negative pairs are generated through random sampling during training.
Evaluation metrics are Recall@$k$ and top-$k$ Normalized Discounted Cumulative Gain (NDCG@$k$), which are evaluated from ground truth items mixed with 100 randomly sampled negative items.
To test the generalizability of user representations, we randomly split the user pool among the training ($80\%$), validation ($10\%$), and test sets ($10\%$).

\noindent\textbf{Public downstream tasks.}
We filter out users and items with fewer than $5$ interactions. Each user's interaction history was listed chronologically. We use item descriptions such as titles, categories, and brands for item information. The maximum token length of item text is set to $512$.
Following previous works~\citep{kang2018self, sun2019bert4rec, hou2022towards}, we adopt the leave-one-out strategy, i.e., next item recommendation task. The last item, second last item, and other items are used as the test, validation, and training data respectively. The Recall@$k$ and NDCG@$k$ are computed by ranking the ground-truth item among all the other items. 
% Since these tasks are next-item recommendation tasks, computing prediction score is different from \cref{eq:4}. We compute the softmax probability over the candidate set composed of the positive and sampled negative items as in conventional sequential recommendation.

\subsection{Baselines}
We compare our models against six strong baselines. Behavior Sequence Transformer (BST)~\citep{chen2019behavior} and LightGCN~\citep{he2020lightgcn} are primarily used baselines in various tasks and domains. 
To reflect the recent trend of user modeling research,
which adopts pretrain-then-transfer strategies, we employ several models from these lines of work. UserBERT~\citep{10.1145/3477495.3531810} and UniSRec~\citep{hou2022towards} pretrain self-supervision objectives with language embeddings and then finetune the model to downstream tasks.
The most comparable unified user models to our methods are M6-Rec~\citep{cui2022m6} and CLUE~\citep{shin2021scaling}. These two methods treat user history as plain text and construct a universal encoder that can be adapted to any domain and task. Note that all the pretrain-then-transfer models, excluding CLUE, utilize webtext language models.
Please see \cref{appendix_2} for more details of baselines.

\setlength{\tabcolsep}{4.3pt}
\ctable[
% 	captionskip = 0pt,
    caption = {Results on the various downstream tasks from in-house and public datasets. The best and second-best results are denoted in bold and underlined, respectively. “Improv.” indicates the relative improvement of our methods over the best baselines.},
    botcap,
    label = tab:eval,
    pos=ht,
%    width=tabularx,
%    figure,
    star,
%  	doinside=\normalsize
    %   doinside=\small
    %  doinside = \footnotesize
    doinside = \scriptsize
    % doinside = \tiny
]{cccccccccccc}{
}{
\toprule 
\multirow{2}{*}{Downstream tasks}
& \multirow{2}{*}{Metrics} 
& \multicolumn{4}{c}{Only trained on task-specific data} 
& \multicolumn{5}{c}{Use additional task-agnostic data}  
& \multirow{2}{*}{Improv.} 
\\
\cmidrule(l){3-6}  
\cmidrule(l){7-11}  
&   & BST  & LightGCN & \name$_{\tiny\text{-lm}}$ & \name & UserBERT & UniSRec & CLUE & M6Rec & \name$_{\tiny\text{+agnostic}}$  & 
\\ 
\midrule[0.52pt]
\midrule[0.52pt]
\multirow{2}{*}{\makecell{OBS}} 
& Recall@10  & 0.4675 & 0.4628  & 0.4654  & \underline{0.4867}  & 0.4600  & 0.4745   & 0.4580 & 0.4615  & \textbf{0.5060} & $+6.6\%$ \\
& NDCG@10    & 0.2780 & 0.2759  & 0.2762  & \underline{0.2940}  & 0.2738  & 0.2825   & 0.2691 & 0.2754  & \textbf{0.3048} & $+7.9\%$ \\
\midrule[0.2pt]
\multirow{2}{*}{\makecell{OTA}} 
& Recall@10  & 0.7160 & 0.7277  & 0.7190  & \underline{0.7428}  & 0.7199  & 0.7186  & 0.7225 & 0.7314   & \textbf{0.7458} & $+2.0\%$ \\
& NDCG@10    & 0.4092 & 0.4235  & 0.4151  & \underline{0.4407}  & 0.4145  & 0.4144  & 0.4219 & 0.4306   & \textbf{0.4431} & $+2.9\%$ \\
\midrule[0.2pt]
\multirow{2}{*}{\makecell{ECOMM}} 
& Recall@10  & 0.6611 & 0.5378  & 0.6667 & \underline{0.7322}   & 0.6934  & 0.6725  & 0.5500 & 0.7093   & \textbf{0.7715} & $+8.8\%$ \\
& NDCG@10    & 0.4846 & 0.4290  & 0.5081 & \underline{0.5637}   & 0.5202  & 0.5079  & 0.4282 & 0.5090   & \textbf{0.6009} & $+15.5\%$ \\
\midrule[1.0pt]
\multirow{2}{*}{\makecell{Scientific}} 
& Recall@10  & 0.0625 & 0.0540 & 0.0951  & \underline{0.1264}   & 0.1055  & 0.1188  & 0.0894 & 0.0945   & \textbf{0.1283} & $+8.0\%$ \\
& NDCG@10    & 0.0323 & 0.0276 & 0.0428  & \underline{0.0695}   & 0.0457  & 0.0641  & 0.0393 & 0.0413   & \textbf{0.0701} & $+9.4\%$ \\
\midrule[0.2pt]
\multirow{2}{*}{\makecell{Pantry}} 
& Recall@10  & 0.0388 & 0.0402 & 0.0626  & \textbf{0.0692}   & 0.0630  & 0.0636  & 0.0602 & 0.0645      & \underline{0.0683} & $+7.3\%$ \\
& NDCG@10    & 0.0203 & 0.0195 & 0.0298  & \textbf{0.0343}   & 0.0312  & 0.0306  & 0.0288 & 0.0324      & \underline{0.0330} & $+5.7\%$ \\
\midrule[0.2pt]
\multirow{2}{*}{\makecell{Online Retail}} 
& Recall@10  & 0.1460 & 0.1322 & 0.1373  & \underline{0.1475}   & 0.1438  & 0.1449  & 0.1258 & 0.1458   & \textbf{0.1502} & $+3.0\%$ \\
& NDCG@10    & 0.0685 & 0.0608 & 0.0659  & \underline{0.0718}   & 0.0654  & 0.0677  & 0.0585 & 0.0702   & \textbf{0.0732} & $+4.3\%$ \\
\bottomrule 
}

\setlength{\tabcolsep}{12pt}
\ctable[
% 	captionskip = 0pt,
    caption = {Performance on the OBS task while varying the ratio of the leveraged task-specific and task-agnostic data for language modeling. We set LMREC's task-specific data size as $100\%$ and vary the task-specific and task-agnostic data ratio.},
    botcap,
    label=tab:corpus,
    pos=t,
%    width=tabularx,
%    figure,
    % star,
%  	doinside=\normalsize
    %   doinside=\small
    %  doinside = \footnotesize
    doinside = \scriptsize
    % doinside = \tiny
]{ccc}{
}{
\toprule 
\multirow{2}{*}{Method} 
& \multicolumn{2}{c}{OBS} 
\\
\cmidrule(l){2-3}
& Recall@10 & NDCG@10 \\
\midrule[0.52pt]
\midrule[0.52pt]
\name$_{\tiny\text{+agnostic}}$ (\texttt{0\%} : \texttt{100\%}) & 0.4703 & 0.2805  \\
\name$_{\tiny\text{+agnostic}}$ (\texttt{30\%} : \texttt{70\%}) & 0.4811 & 0.2932  \\
\name$_{\tiny\text{+agnostic}}$ (\texttt{50\%} : \texttt{50\%}) & 0.4905 & 0.2991  \\
\name$_{\tiny\text{+agnostic}}$ (\texttt{70\%} : \texttt{30\%}) & 0.4917 & 0.3003  \\
\name~                      (\texttt{100\%} : \texttt{0\%}) & 0.4867 & 0.2940  \\
\bottomrule
}

\setlength{\tabcolsep}{3pt}
\ctable[
% 	captionskip = 0pt,
    caption = {Inference time and trainable weight comparison of the downstream models measured from the OBS task. We calculate the inference time of a single batch on A100 GPU.},
    botcap,
    label = tab:resource,
    pos=t,
%    width=tabularx,
%    figure,
    % star,
%  	doinside=\normalsize
      doinside=\small
    %  doinside = \footnotesize
    % doinside = \scriptsize
    % doinside = \tiny
]{cccccc}{
\tnote[$\dagger$] {All the models, excluding LightGCN and CLUE.}}{
\toprule
Models & Inputs & Speedup & Parameters  \\
\midrule
Transformer$^\dagger$   & User history logs    & 1 & 125M  \\
LightGCN              & User history logs    & $\times$34 & 2M  \\
\name$^{\tiny\text{TL}}$   & Pretrained user repr. & $\times$157 & 1.2M  \\
\bottomrule
}

\setlength{\tabcolsep}{1.9pt}
\ctable[
% 	captionskip = 0pt,
    caption = {Task-agnostic transfer learning results on in-house datasets. All the models are pretrained with datasets that are not the target task. For example, the models are first pretrained with OBS and OTA and then transferred to the ECOMM (target task). The “Task-specific feature” stand for the models that use task-specific user data to produce user embedding, while the “Task-agnostic feature” stands for user embedding from task-agnostic data, including Search, E-commerce, SNS, and News. The combination of them is denoted as "Combine".},
    botcap,
    label = tab:tl-eval,
    pos=t,
%    width=tabularx,
%    figure,
    star,
%  	doinside=\normalsize
    %   doinside=\small
    %  doinside = \footnotesize
    doinside = \scriptsize
    % doinside = \tiny
]{ccccccccccccccc}{
}{
\toprule 
\multirow{2}{*}{Downstream tasks}
& \multirow{2}{*}{Metrics} 
& \multicolumn{3}{c}{Task-specific feature} 
& \multicolumn{5}{c}{Task-agnostic feature}  
& \multicolumn{4}{c}{Combine}  
\\
\cmidrule(l){3-5}  
\cmidrule(l){6-10}  
\cmidrule(l){11-14}
&  & \name$^{\tiny\text{TL}}_{\tiny\text{-lm}}$ & \name$^{\tiny\text{TL}}$ & \name$^{\tiny\text{TL}}_{\tiny\text{+agn.}}$ & UniSRec & CLUE & M6Rec & \name$^{\tiny\text{TL}}$ & \name$^{\tiny\text{TL}}_{\tiny\text{+agn.}}$ & UniSRec & CLUE & M6Rec  & \name$^{\tiny\text{TL}}_{\tiny\text{+agn.}}$  \\ 
\midrule[0.52pt]
\midrule[0.52pt]
\multirow{2}{*}{\makecell{OBS}} 
& Recall@10  & 0.3661  & 0.4687  & 0.4861 & 0.5133  & 0.5112  & 0.5451 & 0.4837  & 0.5675  & 0.5397 & 0.5416  & 0.5540  & \textbf{0.5952}  \\
& NDCG@10    & 0.2039  & 0.2792  & 0.2886 & 0.3139  & 0.3204  & 0.3357 & 0.2874  & 0.3514  & 0.3305 & 0.3372  & 0.3391  & \textbf{0.3766}  \\
\midrule[0.2pt]
\multirow{2}{*}{\makecell{OTA}} 
& Recall@10  & 0.5531  & 0.7196  & 0.7375 & 0.7121  & 0.7408  & 0.7285 & 0.7231  & 0.7410  & 0.7201 & 0.7436  & 0.7324  & \textbf{0.7521}  \\
& NDCG@10    & 0.3014  & 0.4119  & 0.4368 & 0.4103  & 0.4414  & 0.4288 & 0.4185  & 0.4421  & 0.4166 & 0.4445  & 0.4297  & \textbf{0.4579}  \\
\midrule[0.2pt]
\multirow{2}{*}{\makecell{ECOMM}} 
& Recall@10  & 0.3202  & 0.7134  & 0.7655 & 0.6068  & 0.5763  & 0.6233 & 0.6273  & 0.6653  & 0.6882 & 0.6370  & 0.7204  & \textbf{0.7803}  \\
& NDCG@10    & 0.3547  & 0.5355  & 0.5878 & 0.4748  & 0.4558  & 0.4810 & 0.4485  & 0.4969  & 0.5204 & 0.4838  & 0.5122  & \textbf{0.6117}  \\
\bottomrule
}

\setlength{\tabcolsep}{9.5pt}
\ctable[
% 	captionskip = 0pt,
    caption = {A Click Through Rate (CTR) and Gross Merchandise Value (GMV) gain on the online product collection task. The user group `new' corresponds to users with no recorded behavior on the service for the past month. We set the GNN model gain as the baseline for the CTR and GMV calculation.},
    botcap,
    label=tab:md,
    pos=t,
%    width=tabularx,
%    figure,
    % star,
%  	doinside=\normalsize
    %   doinside=\small
    %  doinside = \footnotesize
    doinside = \scriptsize
    % doinside = \tiny
]{ccccc}{
}{
\toprule 
\multirow{2}{*}{Method} 
& \multicolumn{2}{c}{CTR} 
& \multicolumn{2}{c}{GMV}  
\\
\cmidrule(l){2-3}
\cmidrule(l){4-5}
& New & Total  & New & Total  \\
\midrule[0.52pt]
\midrule[0.52pt]
GNN         & 1.00    & 1.00   & 1.00   & 1.00  \\
CLUE        & $\times$1.52   & $\times$1.14   & $\times$1.08   & $\times$1.02   \\
\name$^{\tiny\text{TL}}_{\tiny\text{+agnostic}}$  & $\times$1.76   & $\times$1.24   & $\times$1.12    & $\times$1.04   \\
\bottomrule
}
\section{Results} \label{sec:results}
\subsection{Performance on Various Tasks}
\cref{tab:eval} presents the efficacy of our \name~ against baselines. 
Across the six datasets, \name~trained only with the task-specific data achieves state-of-the-art performances compared to all the baselines, even though some methods utilize additional task-agnostic data.
For the in-house datasets, \name~surpasses best performing baseline models by over $1.6\sim3.2\%$ in Recall@$10$. In the public datasets, \name~shows around $5\%$ average improvements compared to baselines. 
Since other pretrain-then-transfer models leverage additional data, we introduce \name$_{\tiny\text{+agnostic}}$, a more robust representation learning method using additional corpus for language modeling. \name$_{\tiny\text{+agnostic}}$~remarkably outperforms the other models in all tasks by a significant margin (see improvement in \cref{tab:eval}).
We further conduct an ablation study on combining task-specific and task-agnostic corpus when the computation resources are limited. \cref{tab:corpus} presents the results. \name$_{\tiny\text{+agnostic}}$ (\texttt{0\%} : \texttt{100\%}), i.e., language modeling on task-agnostic data only, outperforms \name$_{\tiny\text{-lm}}$ in \cref{tab:eval}, but shows the worst performance in \cref{tab:corpus}. Increasing the ratio of used task-specific data delivers performance benefits to some point (\texttt{70\%}). However, leveraging task-specific data solely finally decreases the performance.

Previous research provides a theoretical analysis of why language model pretraining guarantees effective representation learning for downstream tasks~\citep{saunshi2021a, wei2021why}. The additional analysis in~\cref{appendix_3} may support these results.

\subsection{Linear Probe}
We show the effectiveness of the language model pretraining then feature-based transfer strategy (\cref{fig:m2}) across all tasks. Our approach empirically demonstrates the flexible generalizability of the pretrained features. Note that all the baselines, excluding CLUE, are pretrain-then-finetune methods, and the downstream computational cost (\cref{tab:resource}) is much more expensive than the linear probe. 

As shown in \cref{tab:tl-eval}, the linear probe result of \name$^{\tiny\text{TL}}_{\tiny\text{-lm}}$~that are trained only on recommendation tasks shows worst transfer learning performances. Unsurprisingly, a model trained without language modeling cannot guarantee generalizability to other language corpora.
It is worth mentioning that \name$^{\tiny\text{TL}}$, which jointly trains language model and recommendation tasks objectives, shows decent transfer learning capability for downstream tasks. This result provides that incorporating language model pretraining with recommender system profits strong adaptability and generality compared to the recommendation model, even on the linear probe, i.e., not trained on downstream tasks directly.
As previous research~\citep{gururangan-etal-2020-dont, krishna2022downstream} confirmed, it is reasonable to believe that leveraging large quantities of additional data for language model pretraining is strictly more powerful than using small task-specific data. \name$^{\tiny\text{TL}}_{\tiny\text{+agnostic}}$~shows enhanced transferability on linear probe. Comparing results among \cref{tab:eval}, \ref{tab:resource}, and \ref{tab:tl-eval}, we can see that \name$^{\tiny\text{TL}}_{\tiny\text{+agnostic}}$~outperforms other baselines with much fast and easy adaptation.

\begin{figure*}[t]
\begin{center}
\centerline{\includegraphics[width=0.90\textwidth]{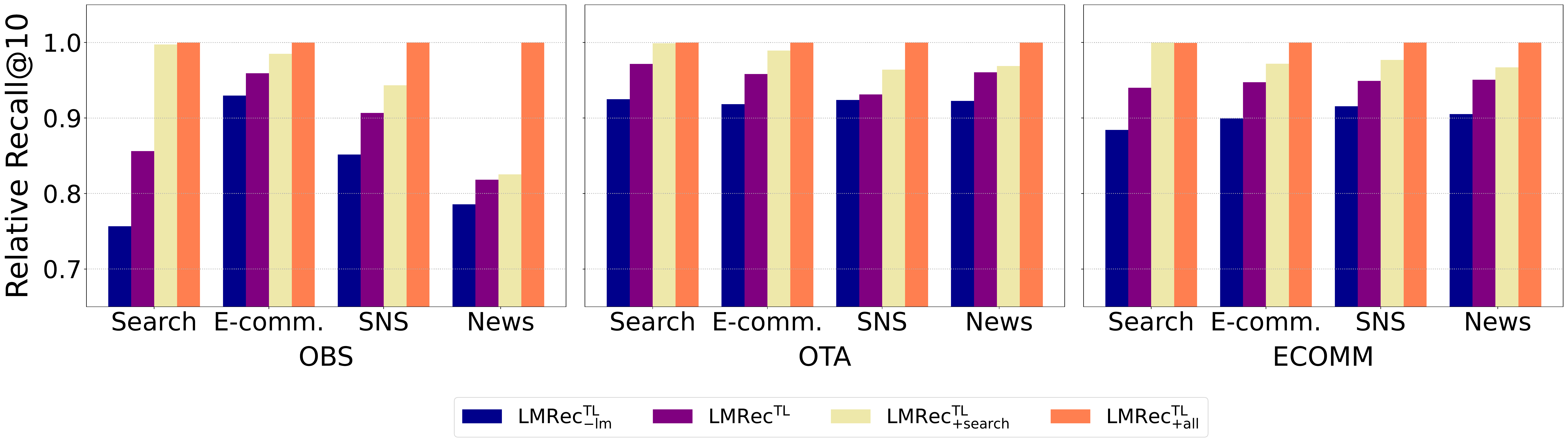}}
\caption{An ablation study on the pretrained user behavior corpora by comparing four types of setup; pretraining on all task-agnostic corpora (\name$^{\tiny\text{TL}}_{\tiny\text{+all}}$), search corpus (\name$^{\tiny\text{TL}}_{\tiny\text{+search}}$), task-specific corpus (\name$^{\tiny\text{TL}}$), and training without language model objectives (\name$^{\tiny\text{TL}}_{\tiny\text{-lm}}$), i.e., train recommendation tasks only. For each pretraining setup, we perform a linear probe on the representation of each task-agnostic user data (x-axis). Linear probe results are normalized across pretrained models for each task-agnostic data. Pretraining protocol follows that of \cref{tab:tl-eval}.}
\label{fig:corpus_abl}
\end{center}
\end{figure*} 
\begin{figure*}[t]
\begin{center}
\centerline{\includegraphics[width=0.90\textwidth]{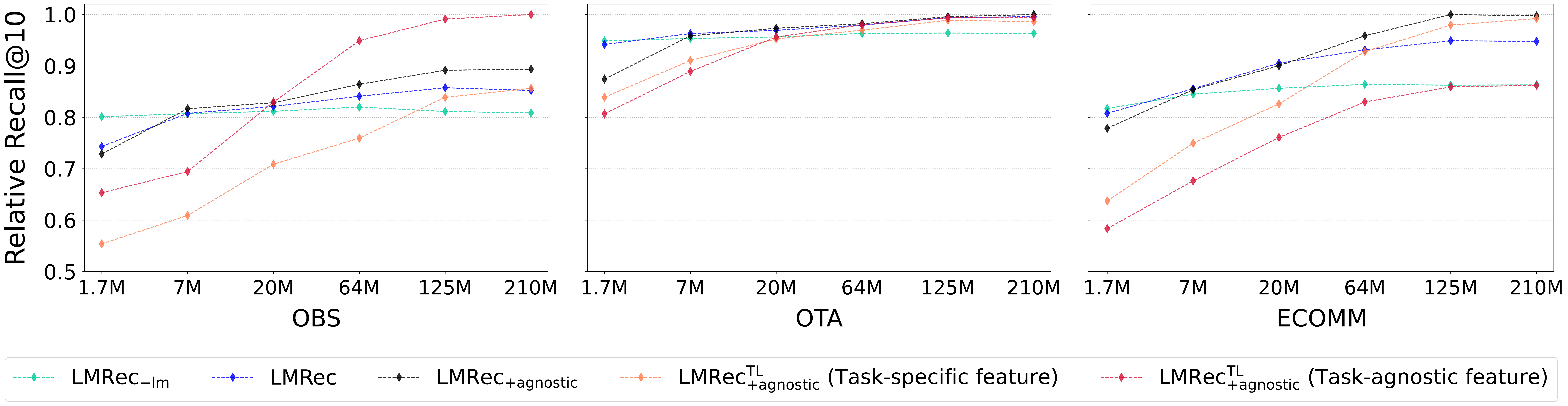}}
\caption{Performance on the downstream tasks according to the size of model parameters ranging from $1.7$ million to $210$ million. The Recall@$10$ is normalized across models for each task.}
\label{fig:size_abl}
\end{center}
\end{figure*} 
\subsection{Virtues of More User Data}
A line of research that studies scaling law in recommender systems argues that parameter growth will not always offer performance improvement and has low return-on-investment (ROI) in resource efficiencies~\citep{ardalani2022understanding, shin2021scaling}. Hence, the data scaling scheme should be treated as a top priority for improving model performances. To verify the efficacy of the data scaling approach, we evaluate our model on downstream tasks by using task-agnostic data as user feature. Results are presented in~\cref{tab:tl-eval}-(Task-agnostic feature/Combine). We simply concatenate task-specific and task-agnostic data to use as inputs for the Combine setup. Most baselines are not adequately reflecting the possibility of using additional user features due to their pretraining methods, but \name$^{\tiny\text{TL}}_{\tiny\text{+agnostic}}$~properly considers the potential of using more user data. It is an enormous benefit to the models seeing that \name$^{\tiny\text{TL}}_{\tiny\text{+agnostic}}$ (Combine) shows outstanding performance by combining all the user data. 
Interestingly, \name$^{\tiny\text{TL}}$, which is trained without task-agnostic data, also achieves state-of-the-art or comparable performances to the baseline models. This result highlights the efficacy of our approach. 

We conducted an online A/B experiment for a product collection recommendation task (see \cref{appendix_4} for more details) on our in-house e-commerce platform for two weeks in August 2022. \cref{tab:md} shows the consistent superiority of our method online. 
For user groups `new', the user representation by \name$^{\tiny\text{TL}}_{\tiny\text{+agnostic}}$~significantly improves CTR and GMV compared to GNN~\citep{jeong2020div2vec}. We conjecture that it may benefit from additional user data from other services, thus contributing to users with no recorded behavior. 

\subsection{Effect of Pretraining Behavior Corpora for Transfer Learning}
We perform ablation studies on the relations between pretraining corpora and using task-agnostic data as user features. 
As shown in \cref{fig:corpus_abl}, the model pretrained with the specific corpus provides general and robust representations of that corpus even on unseen tasks.
Interestingly, tailoring a language model to diverse corpora may bridge the gap between pretraining and task-agnostic corpus domains. For example, even though \name$^{\tiny\text{TL}}_{\tiny\text{+search}}$~leverages only Search corpus for language model pretraining, it consistently outperforms \name$^{\tiny\text{TL}}_{\tiny\text{-lm}}$ and \name$^{\tiny\text{TL}}$ on all the downstream tasks with other task-agnostic features.

As it can be seen in \cref{fig:corpus_abl}, the performance of \name$^{\tiny\text{TL}}_{\tiny\text{-lm}}$ in OBS task is relatively low compared to other tasks. It is due to the strong contribution of task-agnostic features (\cref{tab:tl-eval} and \cref{fig:size_abl}) for the OBS task. In other words, when the task-agnostic features are well-transferable to the target downstream tasks, the performance differences between not pretrained (\name$^{\tiny\text{TL}}_{\tiny\text{-lm}}$) and the rest can be substantial.

\subsection{Effect of Model Size}
Many recent reports in NLP and computer vision have empirically demonstrated the existence of a scaling law, where performance scales strongly with model capacity~\citep{NEURIPS2020_1457c0d6, kaplan2020scaling, zhai2021scaling, bahri2021explaining}. Recently, \citet{shin2021scaling} found the power-law learning curve as a function of model size in recommender systems. \cref{fig:size_abl} shows that scaling up the model leads to a strict performance improvement on the downstream tasks, consistent with the results in the prior works. However, we can also find that models' performances have an upper limit. It is in harmony with the trend in \citet{ardalani2022understanding} that the recommendation performance follows a power law plus a \textit{constant} relationship to the model size, which is an irreducible error on our side.

Note that the performances of \name$^{\tiny\text{TL}}_{\tiny\text{-lm}}$ do not vary according to the model sizes. We conjecture that the model trained without language modeling has no benefits from high model complexity, as its learning capacity is naturally limited.
\section{Related Work}  \label{sec:rw}
Any model that trains a text-based user model to adapt to unseen domains/systems can be viewed as prior work of our research. This line of work has been recently explored since learning text representation has been rapidly developed in the decade. 
In this context, \citet{qiu2021u} and \citet{gu2021exploiting} are the earliest work we are aware of. They train the model through critical word matching in user logs and then finetune models to the downstream tasks. First, the word (item) embeddings are precomputed using pretrained language models (PLMs). The sequence of item embeddings is then passed to the encoder to produce user representations.
Recently, some researchers propose to use behavior history as plain text data~\citep{geng2022recommendation, cui2022m6, hou2022towards, shin2021scaling}.~\citet{hou2022towards} and~\citet{shin2021scaling} introduce a contrastive learning framework on multiple service domains, and perform transfer learning across various downstream tasks. 
Another line of work~\citep{geng2022recommendation, cui2022m6} tries to construct personalized prompts for building versatile framework, i.e., ``Here is the history of \{gender\} \{age\}: \{history from all services\}, The user is now recommended 
a \{item\}''. This approach profits from the methods that utilize language models such as GPT-2~\citep{radford2019language}, T5~\citep{raffel2020exploring}, and M6~\citep{10.1145/3447548.3467206}.
Their PLM-based approach can be generalized to various applications, with the ability to perform zero-shot learning.~\citet{shin2021scaling} is the only work that trained the whole encoder from scratch rather than using PLMs. We refer readers to \citet{liu2023pre} and \citet{yuan2023go} for an overview of this line of work.

A related idea to our work is the training language model on task-specific or task-agnostic corpora. It has been shown to be beneficial in a variety of works~\citep{chronopoulou-etal-2019-embarrassingly, gururangan-etal-2020-dont, lee2020biobert, karouzos-etal-2021-udalm, krishna2022downstream}.~\citet{gururangan-etal-2020-dont} continue pretraining of LM on task-specific data and show it can improve the downstream performances of standard webtext language models.~\citet{krishna2022downstream} point out that the effect of pretraining on standard webtext data may have been overestimated. They show that models trained only on task-specific data comparably perform to existing webtext language models. 
On the one hand, a line of research jointly trains language models on task-specific data during finetuning to avoid catastrophic forgetting~\citep{chronopoulou-etal-2019-embarrassingly, karouzos-etal-2021-udalm}. 
Some of the works above also investigate if the models pretrained on task-agnostic data can be effective for downstream tasks.~\citet{gururangan-etal-2020-dont} and~\citet{lee2020biobert} show domain-adaptive pretraining further improves the performance of pretrained language models. Recently,~\citet{krishna2022downstream} have observed that pretraining on task-agnostic data can provide a significant advantage compared to standard webtext data. These findings give huge insight into our research.

Note that our work aims at extending the potential of language modeling that has been successfully used for diverse applications to recommender systems.

\section{Conclusion}  \label{sec:conc}
Recent works have built text-based user models and demonstrated that the rich nature of text information in any domain or system could be a valuable foundation for user modeling.
Our primary contribution is jointly optimizing the language modeling and recommendation task objectives and successfully tackling a broad spectrum of diverse recommendation tasks, including transfer learning for unseen domains and systems. 
Overall, our analysis sheds remarkable insights on user representation learning through user behavioral corpora.
\section*{Considerations and Limitations}  \label{sec:limit}
LMRec is trained on user behavior text data that are collected from diverse service applications. These datasets are preprocessed to users' behavior sequences as detailed in \cref{fig:overview} and \cref{sec:exp-data}. 
However, in order to improve the quality of user representations, choosing the item information differently for each application may improve the effectiveness.
As such, we can consider domain-specific information for each service rather than using general item information. For example, we may leverage additional domain-specific information such as news topics or categories, names of the press agency, and keywords for the news content rather than using only news titles for the News dataset. This issue is a promising extension for practitioners to successfully apply LMRec to real-world applications.

The types of task-agnostic data will largely affect the performance gains of \name$_{\tiny\text{+agnostic}}$ and \name$^{\tiny\text{TL}}_{\tiny\text{+agnostic}}$. 
We fully utilize four types of task-agnostic data, i.e., Search, E-comm., SNS, and News, and achieve state-of-the-art results. However, this paper does not thoroughly explore their optimized combination or mixing ratio of the corpus due to the heavy computational costs, which most large LM studies suffer from. While prior work shows how the pretraining corpus sources and their combination affect diverse downstream tasks~\citep{raffel2020exploring, gururangan-etal-2020-dont, lee2020biobert, krishna2022downstream, shin2022effect}, there still remain limitations in finding the generic relation between downstream performance and corpus properties; measuring the effect of the pretraining corpus on the downstream task is still underexplored. We point out that more careful study is left for future research.

Regarding reproducibility, it is difficult to open our in-house data due to legal issues caused by privacy and user agreement. Therefore, we tried our best to validate the efficacy of our LMRec with the experiments on benchmark datasets in addition to in-house data.

\section*{Acknowledgements}
All authors thank NAVER Smart Machine Learning (NSML) platform team~\citep{sung2017nsml, kim2018nsml} for their critical work on the software and hardware infrastructure on which all the experiments were performed.
\bibliography{anthology}

\begin{thebibliography}{59}
\expandafter\ifx\csname natexlab\endcsname\relax\def\natexlab#1{#1}\fi

\bibitem[{Ardalani et~al.(2022)Ardalani, Wu, Chen, Bhushanam, and
  Aziz}]{ardalani2022understanding}
Newsha Ardalani, Carole-Jean Wu, Zeliang Chen, Bhargav Bhushanam, and Adnan
  Aziz. 2022.
\newblock Understanding scaling laws for recommendation models.
\newblock \emph{arXiv preprint arXiv:2208.08489}.

\bibitem[{Bahri et~al.(2021)Bahri, Dyer, Kaplan, Lee, and
  Sharma}]{bahri2021explaining}
Yasaman Bahri, Ethan Dyer, Jared Kaplan, Jaehoon Lee, and Utkarsh Sharma. 2021.
\newblock Explaining neural scaling laws.
\newblock \emph{arXiv preprint arXiv:2102.06701}.

\bibitem[{Borsos et~al.(2022)Borsos, Marinier, Vincent, Kharitonov, Pietquin,
  Sharifi, Teboul, Grangier, Tagliasacchi, and Zeghidour}]{borsos2022audiolm}
Zal{\'a}n Borsos, Rapha{\"e}l Marinier, Damien Vincent, Eugene Kharitonov,
  Olivier Pietquin, Matt Sharifi, Olivier Teboul, David Grangier, Marco
  Tagliasacchi, and Neil Zeghidour. 2022.
\newblock Audiolm: a language modeling approach to audio generation.
\newblock \emph{arXiv preprint arXiv:2209.03143}.

\bibitem[{Brown et~al.(2020)Brown, Mann, Ryder, Subbiah, Kaplan
  et~al.}]{NEURIPS2020_1457c0d6}
Tom Brown, Benjamin Mann, Nick Ryder, Melanie Subbiah, Jared~D Kaplan, et~al.
  2020.
\newblock Language models are few-shot learners.
\newblock In \emph{Advances in Neural Information Processing Systems},
  volume~33, pages 1877--1901. Curran Associates, Inc.

\bibitem[{Chen et~al.(2020)Chen, Radford, Child, Wu, Jun, Luan, and
  Sutskever}]{chen2020generative}
Mark Chen, Alec Radford, Rewon Child, Jeffrey Wu, Heewoo Jun, David Luan, and
  Ilya Sutskever. 2020.
\newblock Generative pretraining from pixels.
\newblock In \emph{International conference on machine learning}, pages
  1691--1703. PMLR.

\bibitem[{Chen et~al.(2021)Chen, Tworek, Jun, Yuan, Pinto, Kaplan, Edwards,
  Burda, Joseph, Brockman et~al.}]{chen2021evaluating}
Mark Chen, Jerry Tworek, Heewoo Jun, Qiming Yuan, Henrique Ponde de~Oliveira
  Pinto, Jared Kaplan, Harri Edwards, Yuri Burda, Nicholas Joseph, Greg
  Brockman, et~al. 2021.
\newblock Evaluating large language models trained on code.
\newblock \emph{arXiv preprint arXiv:2107.03374}.

\bibitem[{Chen et~al.(2019)Chen, Zhao et~al.}]{chen2019behavior}
Qiwei Chen, Huan Zhao, et~al. 2019.
\newblock Behavior sequence transformer for e-commerce recommendation in
  alibaba.
\newblock In \emph{Proceedings of the 1st International Workshop on Deep
  Learning Practice for High-Dimensional Sparse Data}, pages 1--4.

\bibitem[{Chen et~al.(2022)Chen, Hsieh, and Gong}]{chen2022when}
Xiangning Chen, Cho-Jui Hsieh, and Boqing Gong. 2022.
\newblock \href {https://openreview.net/forum?id=LtKcMgGOeLt} {When vision
  transformers outperform resnets without pre-training or strong data
  augmentations}.
\newblock In \emph{International Conference on Learning Representations}.

\bibitem[{Chronopoulou et~al.(2019)Chronopoulou, Baziotis, and
  Potamianos}]{chronopoulou-etal-2019-embarrassingly}
Alexandra Chronopoulou, Christos Baziotis, and Alexandros Potamianos. 2019.
\newblock \href {https://doi.org/10.18653/v1/N19-1213} {An embarrassingly
  simple approach for transfer learning from pretrained language models}.
\newblock In \emph{Proceedings of the 2019 Conference of the North {A}merican
  Chapter of the Association for Computational Linguistics: Human Language
  Technologies, Volume 1 (Long and Short Papers)}, pages 2089--2095,
  Minneapolis, Minnesota. Association for Computational Linguistics.

\bibitem[{Cui et~al.(2022)Cui, Ma, Zhou, Zhou, and Yang}]{cui2022m6}
Zeyu Cui, Jianxin Ma, Chang Zhou, Jingren Zhou, and Hongxia Yang. 2022.
\newblock M6-rec: Generative pretrained language models are open-ended
  recommender systems.
\newblock \emph{arXiv preprint arXiv:2205.08084}.

\bibitem[{Foret et~al.(2021)Foret, Kleiner, Mobahi, and
  Neyshabur}]{foret2021sharpnessaware}
Pierre Foret, Ariel Kleiner, Hossein Mobahi, and Behnam Neyshabur. 2021.
\newblock \href {https://openreview.net/forum?id=6Tm1mposlrM} {Sharpness-aware
  minimization for efficiently improving generalization}.
\newblock In \emph{International Conference on Learning Representations}.

\bibitem[{Geng et~al.(2022)Geng, Liu, Fu, Ge, and
  Zhang}]{geng2022recommendation}
Shijie Geng, Shuchang Liu, Zuohui Fu, Yingqiang Ge, and Yongfeng Zhang. 2022.
\newblock \href {https://doi.org/10.1145/3523227.3546767} {Recommendation as
  language processing (rlp): A unified pretrain, personalized prompt \& predict
  paradigm (p5)}.
\newblock In \emph{Proceedings of the 16th ACM Conference on Recommender
  Systems}, RecSys '22, page 299–315, New York, NY, USA. Association for
  Computing Machinery.

\bibitem[{Gu et~al.(2021)Gu, Wang, Sun, Ye, Xu, Chen, and
  Zhang}]{gu2021exploiting}
Jie Gu, Feng Wang, Qinghui Sun, Zhiquan Ye, Xiaoxiao Xu, Jingmin Chen, and Jun
  Zhang. 2021.
\newblock Exploiting behavioral consistence for universal user representation.
\newblock In \emph{Proceedings of the AAAI Conference on Artificial
  Intelligence}, volume~35, pages 4063--4071.

\bibitem[{Gururangan et~al.(2020)Gururangan, Marasovi{\'c}, Swayamdipta, Lo,
  Beltagy, Downey, and Smith}]{gururangan-etal-2020-dont}
Suchin Gururangan, Ana Marasovi{\'c}, Swabha Swayamdipta, Kyle Lo, Iz~Beltagy,
  Doug Downey, and Noah~A. Smith. 2020.
\newblock \href {https://doi.org/10.18653/v1/2020.acl-main.740} {Don{'}t stop
  pretraining: Adapt language models to domains and tasks}.
\newblock In \emph{Proceedings of the 58th Annual Meeting of the Association
  for Computational Linguistics}, pages 8342--8360, Online. Association for
  Computational Linguistics.

\bibitem[{He et~al.(2020)He, Deng, Wang, Li, Zhang, and Wang}]{he2020lightgcn}
Xiangnan He, Kuan Deng, Xiang Wang, Yan Li, YongDong Zhang, and Meng Wang.
  2020.
\newblock \href {https://doi.org/10.1145/3397271.3401063} {Lightgcn:
  Simplifying and powering graph convolution network for recommendation}.
\newblock In \emph{Proceedings of the 43rd International ACM SIGIR Conference
  on Research and Development in Information Retrieval}, SIGIR '20, page
  639–648, New York, NY, USA. Association for Computing Machinery.

\bibitem[{Hou et~al.(2022)Hou, Mu, Zhao, Li, Ding, and Wen}]{hou2022towards}
Yupeng Hou, Shanlei Mu, Wayne~Xin Zhao, Yaliang Li, Bolin Ding, and Ji-Rong
  Wen. 2022.
\newblock \href {https://doi.org/10.1145/3534678.3539381} {Towards universal
  sequence representation learning for recommender systems}.
\newblock In \emph{Proceedings of the 28th ACM SIGKDD Conference on Knowledge
  Discovery and Data Mining}, KDD '22, page 585–593, New York, NY, USA.
  Association for Computing Machinery.

\bibitem[{Jeong et~al.(2020)Jeong, Yun, Keam et~al.}]{jeong2020div2vec}
Jisu Jeong, Jeong-Min Yun, Hongi Keam, et~al. 2020.
\newblock div2vec: Diversity-emphasized node embedding.
\newblock In \emph{ImpactRS Workshop at Recsys 2020}.

\bibitem[{Kang and McAuley(2018)}]{kang2018self}
Wang-Cheng Kang and Julian McAuley. 2018.
\newblock Self-attentive sequential recommendation.
\newblock In \emph{2018 IEEE international conference on data mining (ICDM)},
  pages 197--206. IEEE.

\bibitem[{Kaplan et~al.(2020)Kaplan, McCandlish, Henighan, Brown
  et~al.}]{kaplan2020scaling}
Jared Kaplan, Sam McCandlish, Tom Henighan, Tom~B Brown, et~al. 2020.
\newblock Scaling laws for neural language models.
\newblock \emph{arXiv preprint arXiv:2001.08361}.

\bibitem[{Karouzos et~al.(2021)Karouzos, Paraskevopoulos, and
  Potamianos}]{karouzos-etal-2021-udalm}
Constantinos Karouzos, Georgios Paraskevopoulos, and Alexandros Potamianos.
  2021.
\newblock \href {https://doi.org/10.18653/v1/2021.naacl-main.203} {{UDALM}:
  Unsupervised domain adaptation through language modeling}.
\newblock In \emph{Proceedings of the NAACL-HLT}, pages 2579--2590, Online.
  Association for Computational Linguistics.

\bibitem[{Kim et~al.(2018)Kim, Kim, Seo, Kim, Park, Park, Jo, Kim, Yang, Kim
  et~al.}]{kim2018nsml}
Hanjoo Kim, Minkyu Kim, Dongjoo Seo, Jinwoong Kim, Heungseok Park, Soeun Park,
  Hyunwoo Jo, KyungHyun Kim, Youngil Yang, Youngkwan Kim, et~al. 2018.
\newblock Nsml: Meet the mlaas platform with a real-world case study.
\newblock \emph{arXiv preprint arXiv:1810.09957}.

\bibitem[{Kipf and Welling(2017)}]{kipf2016semi}
Thomas~N. Kipf and Max Welling. 2017.
\newblock \href {https://openreview.net/forum?id=SJU4ayYgl} {Semi-supervised
  classification with graph convolutional networks}.
\newblock In \emph{International Conference on Learning Representations}.

\bibitem[{Krishna et~al.(2022)Krishna, Garg, Bigham, and
  Lipton}]{krishna2022downstream}
Kundan Krishna, Saurabh Garg, Jeffrey~P Bigham, and Zachary~C Lipton. 2022.
\newblock Downstream datasets make surprisingly good pretraining corpora.
\newblock \emph{arXiv preprint arXiv:2209.14389}.

\bibitem[{Lee et~al.(2020)Lee, Yoon, Kim, Kim, Kim, So, and
  Kang}]{lee2020biobert}
Jinhyuk Lee, Wonjin Yoon, Sungdong Kim, Donghyeon Kim, Sunkyu Kim, Chan~Ho So,
  and Jaewoo Kang. 2020.
\newblock Biobert: a pre-trained biomedical language representation model for
  biomedical text mining.
\newblock \emph{Bioinformatics}, 36(4):1234--1240.

\bibitem[{Li et~al.(2018)Li, Xu, Taylor, Studer, and
  Goldstein}]{li2018visualizing}
Hao Li, Zheng Xu, Gavin Taylor, Christoph Studer, and Tom Goldstein. 2018.
\newblock Visualizing the loss landscape of neural nets.
\newblock \emph{Advances in neural information processing systems}, 31.

\bibitem[{Lin et~al.(2021)Lin, Men, Yang, Zhou, Zhang, Wang, Zhou, Tang, and
  Yang}]{10.1145/3447548.3467206}
Junyang Lin, Rui Men, An~Yang, Chang Zhou, Yichang Zhang, Peng Wang, Jingren
  Zhou, Jie Tang, and Hongxia Yang. 2021.
\newblock \href {https://doi.org/10.1145/3447548.3467206} {M6:
  Multi-modality-to-multi-modality multitask mega-transformer for unified
  pretraining}.
\newblock In \emph{Proceedings of the 27th ACM SIGKDD Conference on Knowledge
  Discovery and Data Mining}, KDD '21, page 3251–3261, New York, NY, USA.
  Association for Computing Machinery.

\bibitem[{Liu et~al.(2023)Liu, Zhang, and Gulla}]{liu2023pre}
Peng Liu, Lemei Zhang, and Jon~Atle Gulla. 2023.
\newblock Pre-train, prompt and recommendation: A comprehensive survey of
  language modelling paradigm adaptations in recommender systems.
\newblock \emph{arXiv preprint arXiv:2302.03735}.

\bibitem[{Liu et~al.(2019)Liu, Ott, Goyal, Du, Joshi, Chen, Levy, Lewis,
  Zettlemoyer, and Stoyanov}]{liu2019roberta}
Yinhan Liu, Myle Ott, Naman Goyal, Jingfei Du, Mandar Joshi, Danqi Chen, Omer
  Levy, Mike Lewis, Luke Zettlemoyer, and Veselin Stoyanov. 2019.
\newblock Roberta: A robustly optimized bert pretraining approach.
\newblock \emph{arXiv preprint arXiv:1907.11692}.

\bibitem[{Loshchilov and Hutter(2017)}]{loshchilov2016sgdr}
Ilya Loshchilov and Frank Hutter. 2017.
\newblock Sgdr: Stochastic gradient descent with warm restarts.
\newblock In \emph{International Conference on Learning Representations}.

\bibitem[{Loshchilov and Hutter(2019)}]{loshchilov2018decoupled}
Ilya Loshchilov and Frank Hutter. 2019.
\newblock \href {https://openreview.net/forum?id=Bkg6RiCqY7} {Decoupled weight
  decay regularization}.
\newblock In \emph{International Conference on Learning Representations}.

\bibitem[{Lu et~al.(2020)Lu, Jiao, and Zhang}]{10.1145/3340531.3412747}
Wenhao Lu, Jian Jiao, and Ruofei Zhang. 2020.
\newblock \href {https://doi.org/10.1145/3340531.3412747} {Twinbert: Distilling
  knowledge to twin-structured compressed bert models for large-scale
  retrieval}.
\newblock In \emph{Proceedings of the 29th ACM International Conference on
  Information \& Knowledge Management}, CIKM '20, page 2645–2652, New York,
  NY, USA. Association for Computing Machinery.

\bibitem[{Man et~al.(2017)Man, Shen, Jin, and Cheng}]{man2017cross}
Tong Man, Huawei Shen, Xiaolong Jin, and Xueqi Cheng. 2017.
\newblock Cross-domain recommendation: An embedding and mapping approach.
\newblock In \emph{IJCAI}, volume~17, pages 2464--2470.

\bibitem[{Micikevicius et~al.(2018)Micikevicius, Narang, Alben, Diamos, Elsen,
  Garcia, Ginsburg, Houston, Kuchaiev, Venkatesh, and
  Wu}]{micikevicius2018mixed}
Paulius Micikevicius, Sharan Narang, Jonah Alben, Gregory Diamos, Erich Elsen,
  David Garcia, Boris Ginsburg, Michael Houston, Oleksii Kuchaiev, Ganesh
  Venkatesh, and Hao Wu. 2018.
\newblock \href {https://openreview.net/forum?id=r1gs9JgRZ} {Mixed precision
  training}.
\newblock In \emph{International Conference on Learning Representations}.

\bibitem[{Neelakantan et~al.(2022)Neelakantan, Xu, Puri, Radford, Han, Tworek,
  Yuan, Tezak, Kim, Hallacy et~al.}]{neelakantan2022text}
Arvind Neelakantan, Tao Xu, Raul Puri, Alec Radford, Jesse~Michael Han, Jerry
  Tworek, Qiming Yuan, Nikolas Tezak, Jong~Wook Kim, Chris Hallacy, et~al.
  2022.
\newblock Text and code embeddings by contrastive pre-training.
\newblock \emph{arXiv preprint arXiv:2201.10005}.

\bibitem[{Ni et~al.(2019)Ni, Li, and McAuley}]{ni2019justifying}
Jianmo Ni, Jiacheng Li, and Julian McAuley. 2019.
\newblock Justifying recommendations using distantly-labeled reviews and
  fine-grained aspects.
\newblock In \emph{Proceedings of the EMNLP-IJCNLP}, pages 188--197.

\bibitem[{Park and Kim(2022)}]{park2022how}
Namuk Park and Songkuk Kim. 2022.
\newblock \href {https://openreview.net/forum?id=D78Go4hVcxO} {How do vision
  transformers work?}
\newblock In \emph{International Conference on Learning Representations}.

\bibitem[{Pascanu et~al.(2013)Pascanu, Mikolov, and
  Bengio}]{pascanu2013difficulty}
Razvan Pascanu, Tomas Mikolov, and Yoshua Bengio. 2013.
\newblock On the difficulty of training recurrent neural networks.
\newblock In \emph{International Conference on Machine Learning}.

\bibitem[{Paszke et~al.(2019)Paszke, Gross, Massa, Lerer, Bradbury, Chanan,
  Killeen, Lin, Gimelshein, Antiga, Desmaison, Kopf, Yang, DeVito, Raison,
  Tejani, Chilamkurthy, Steiner, Fang, Bai, and
  Chintala}]{NEURIPS2019_bdbca288}
Adam Paszke, Sam Gross, Francisco Massa, Adam Lerer, James Bradbury, Gregory
  Chanan, Trevor Killeen, Zeming Lin, Natalia Gimelshein, Luca Antiga, Alban
  Desmaison, Andreas Kopf, Edward Yang, Zachary DeVito, Martin Raison, Alykhan
  Tejani, Sasank Chilamkurthy, Benoit Steiner, Lu~Fang, Junjie Bai, and Soumith
  Chintala. 2019.
\newblock \href
  {https://proceedings.neurips.cc/paper/2019/file/bdbca288fee7f92f2bfa9f7012727740-Paper.pdf}
  {Pytorch: An imperative style, high-performance deep learning library}.
\newblock In \emph{Advances in Neural Information Processing Systems},
  volume~32. Curran Associates, Inc.

\bibitem[{Pedregosa et~al.(2011)Pedregosa, Varoquaux, Gramfort, Michel,
  Thirion, Grisel, Blondel, Prettenhofer, Weiss, Dubourg, Vanderplas, Passos,
  Cournapeau, Brucher, Perrot, and Duchesnay}]{scikit-learn}
F.~Pedregosa, G.~Varoquaux, A.~Gramfort, V.~Michel, B.~Thirion, O.~Grisel,
  M.~Blondel, P.~Prettenhofer, R.~Weiss, V.~Dubourg, J.~Vanderplas, A.~Passos,
  D.~Cournapeau, M.~Brucher, M.~Perrot, and E.~Duchesnay. 2011.
\newblock Scikit-learn: Machine learning in {P}ython.
\newblock \emph{Journal of Machine Learning Research}, 12:2825--2830.

\bibitem[{Qiu et~al.(2021)Qiu, Wu, Gao, and Fan}]{qiu2021u}
Zhaopeng Qiu, Xian Wu, Jingyue Gao, and Wei Fan. 2021.
\newblock U-bert: Pre-training user representations for improved
  recommendation.
\newblock In \emph{Proceedings of the AAAI Conference on Artificial
  Intelligence}, volume~35, pages 4320--4327.

\bibitem[{Radford et~al.(2019)Radford, Wu, Child, Luan, Amodei, Sutskever
  et~al.}]{radford2019language}
Alec Radford, Jeffrey Wu, Rewon Child, David Luan, Dario Amodei, Ilya
  Sutskever, et~al. 2019.
\newblock Language models are unsupervised multitask learners.
\newblock \emph{OpenAI blog}, 1(8):9.

\bibitem[{Raffel et~al.(2020)Raffel, Shazeer, Roberts, Lee, Narang, Matena,
  Zhou, Li, Liu et~al.}]{raffel2020exploring}
Colin Raffel, Noam Shazeer, Adam Roberts, Katherine Lee, Sharan Narang, Michael
  Matena, Yanqi Zhou, Wei Li, Peter~J Liu, et~al. 2020.
\newblock Exploring the limits of transfer learning with a unified text-to-text
  transformer.
\newblock \emph{J. Mach. Learn. Res.}, 21(140):1--67.

\bibitem[{Rajbhandari et~al.(2020)Rajbhandari, Rasley, Ruwase, and
  He}]{rajbhandari2020zero}
Samyam Rajbhandari, Jeff Rasley, Olatunji Ruwase, and Yuxiong He. 2020.
\newblock Zero: Memory optimizations toward training trillion parameter models.
\newblock In \emph{SC20: International Conference for High Performance
  Computing, Networking, Storage and Analysis}, pages 1--16. IEEE.

\bibitem[{Ramesh et~al.(2021)Ramesh, Pavlov, Goh, Gray, Voss, Radford, Chen,
  and Sutskever}]{ramesh2021zero}
Aditya Ramesh, Mikhail Pavlov, Gabriel Goh, Scott Gray, Chelsea Voss, Alec
  Radford, Mark Chen, and Ilya Sutskever. 2021.
\newblock Zero-shot text-to-image generation.
\newblock In \emph{International Conference on Machine Learning}, pages
  8821--8831. PMLR.

\bibitem[{Saunshi et~al.(2021)Saunshi, Malladi, and Arora}]{saunshi2021a}
Nikunj Saunshi, Sadhika Malladi, and Sanjeev Arora. 2021.
\newblock \href {https://openreview.net/forum?id=vVjIW3sEc1s} {A mathematical
  exploration of why language models help solve downstream tasks}.
\newblock In \emph{International Conference on Learning Representations}.

\bibitem[{Shin et~al.(2021)Shin, Kwak, Kim, Kim, Park, Jeong, and
  Jung}]{shin2021one4all}
Kyuyong Shin, Hanock Kwak, Kyung-Min Kim, Minkyu Kim, Young-Jin Park, Jisu
  Jeong, and Seungjae Jung. 2021.
\newblock One4all user representation for recommender systems in e-commerce.
\newblock \emph{arXiv preprint arXiv:2106.00573}.

\bibitem[{Shin et~al.(2023)Shin, Kwak, Kim, Ramstrom, Jeong, Ha, and
  Kim}]{shin2021scaling}
Kyuyong Shin, Hanock Kwak, Su~Young Kim, Max~Nihlen Ramstrom, Jisu Jeong,
  Jung-Woo Ha, and Kyung-Min Kim. 2023.
\newblock Scaling law for recommendation models: Towards general-purpose user
  representations.
\newblock \emph{Proceedings of the AAAI Conference on Artificial Intelligence}.

\bibitem[{Shin et~al.(2022)Shin, Lee, Ahn, Kim, Kim, Kim, Cho, Lee, Park, Ha
  et~al.}]{shin2022effect}
Seongjin Shin, Sang-Woo Lee, Hwijeen Ahn, Sungdong Kim, HyoungSeok Kim, Boseop
  Kim, Kyunghyun Cho, Gichang Lee, Woomyoung Park, Jung-Woo Ha, et~al. 2022.
\newblock On the effect of pretraining corpora on in-context learning by a
  large-scale language model.
\newblock \emph{Proceedings of the NAACL-HLT}.

\bibitem[{Sun et~al.(2019)Sun, Liu, Wu, Pei, Lin, Ou, and
  Jiang}]{sun2019bert4rec}
Fei Sun, Jun Liu, Jian Wu, Changhua Pei, Xiao Lin, Wenwu Ou, and Peng Jiang.
  2019.
\newblock Bert4rec: Sequential recommendation with bidirectional encoder
  representations from transformer.
\newblock In \emph{Proceedings of the 28th ACM international conference on
  information and knowledge management}, pages 1441--1450.

\bibitem[{Sung et~al.(2017)Sung, Kim, Jo, Yang, Kim, Lausen, Kim, Lee, Kwak, Ha
  et~al.}]{sung2017nsml}
Nako Sung, Minkyu Kim, Hyunwoo Jo, Youngil Yang, Jingwoong Kim, Leonard Lausen,
  Youngkwan Kim, Gayoung Lee, Donghyun Kwak, Jung-Woo Ha, et~al. 2017.
\newblock Nsml: A machine learning platform that enables you to focus on your
  models.
\newblock \emph{arXiv preprint arXiv:1712.05902}.

\bibitem[{Vaswani et~al.(2017)Vaswani, Shazeer, Parmar, Uszkoreit, Jones,
  Gomez, Kaiser, and Polosukhin}]{NIPS2017_3f5ee243}
Ashish Vaswani, Noam Shazeer, Niki Parmar, Jakob Uszkoreit, Llion Jones,
  Aidan~N Gomez, \L~ukasz Kaiser, and Illia Polosukhin. 2017.
\newblock \href
  {https://proceedings.neurips.cc/paper/2017/file/3f5ee243547dee91fbd053c1c4a845aa-Paper.pdf}
  {Attention is all you need}.
\newblock In \emph{Advances in Neural Information Processing Systems},
  volume~30. Curran Associates, Inc.

\bibitem[{Wang et~al.(2020)Wang, Cho, and Gu}]{wang2020neural}
Changhan Wang, Kyunghyun Cho, and Jiatao Gu. 2020.
\newblock Neural machine translation with byte-level subwords.
\newblock In \emph{AAAI}.

\bibitem[{Wei et~al.(2021)Wei, Xie, and Ma}]{wei2021why}
Colin Wei, Sang~Michael Xie, and Tengyu Ma. 2021.
\newblock \href {https://openreview.net/forum?id=MDMV2SxCboX} {Why do
  pretrained language models help in downstream tasks? an analysis of head and
  prompt tuning}.
\newblock In \emph{Advances in Neural Information Processing Systems}.

\bibitem[{Wu et~al.(2022)Wu, Wu, Qi, and Huang}]{10.1145/3477495.3531810}
Chuhan Wu, Fangzhao Wu, Tao Qi, and Yongfeng Huang. 2022.
\newblock \href {https://doi.org/10.1145/3477495.3531810} {Userbert:
  Pre-training user model with contrastive self-supervision}.
\newblock In \emph{Proceedings of the 45th International ACM SIGIR Conference
  on Research and Development in Information Retrieval}, SIGIR '22, page
  2087–2092, New York, NY, USA. Association for Computing Machinery.

\bibitem[{Yao et~al.(2020)Yao, Gholami, Keutzer, and
  Mahoney}]{yao2020pyhessian}
Zhewei Yao, Amir Gholami, Kurt Keutzer, and Michael~W Mahoney. 2020.
\newblock Pyhessian: Neural networks through the lens of the hessian.
\newblock In \emph{2020 IEEE international conference on big data (Big data)},
  pages 581--590. IEEE.

\bibitem[{Yuan et~al.(2019)Yuan, Yao, and Benatallah}]{10.5555/3367471.3367629}
Feng Yuan, Lina Yao, and Boualem Benatallah. 2019.
\newblock Darec: Deep domain adaptation for cross-domain recommendation via
  transferring rating patterns.
\newblock In \emph{Proceedings of the 28th International Joint Conference on
  Artificial Intelligence}, IJCAI'19, page 4227–4233. AAAI Press.

\bibitem[{Yuan et~al.(2023)Yuan, Yuan, Song, Li, Fu, Yang, Pan, and
  Ni}]{yuan2023go}
Zheng Yuan, Fajie Yuan, Yu~Song, Youhua Li, Junchen Fu, Fei Yang, Yunzhu Pan,
  and Yongxin Ni. 2023.
\newblock Where to go next for recommender systems? id-vs. modality-based
  recommender models revisited.
\newblock \emph{arXiv preprint arXiv:2303.13835}.

\bibitem[{Zhai et~al.(2021)Zhai, Kolesnikov, Houlsby, and
  Beyer}]{zhai2021scaling}
Xiaohua Zhai, Alexander Kolesnikov, Neil Houlsby, and Lucas Beyer. 2021.
\newblock Scaling vision transformers.
\newblock \emph{arXiv preprint arXiv:2106.04560}.

\bibitem[{Zhu et~al.(2022)Zhu, Tang, Liu, Zhuang, Xie, Zhang, Lin, and
  He}]{zhu2022personalized}
Yongchun Zhu, Zhenwei Tang, Yudan Liu, Fuzhen Zhuang, Ruobing Xie, Xu~Zhang,
  Leyu Lin, and Qing He. 2022.
\newblock Personalized transfer of user preferences for cross-domain
  recommendation.
\newblock In \emph{Proceedings of the Fifteenth ACM International Conference on
  Web Search and Data Mining}, pages 1507--1515.

\end{thebibliography}
\bibliographystyle{acl_natbib}
\clearpage
\setlength{\tabcolsep}{8pt}
\ctable[
% 	captionskip = 0pt,
    caption = {Architectures and hyperparameters of the models.},
    botcap,
    label = tab:size-param,
    pos=ht,
%    width=tabularx,
%    figure,
     star,
%  	doinside=\normalsize
      doinside=\small
    %  doinside = \footnotesize
    % doinside = \scriptsize
    % doinside = \tiny
]{ccccccccc}{
}{
\toprule
Model Size & $n_{layers}$ & $d_{emb}$ & $n_{heads}$ & $d_{ffn}$ & $\lambda$ & Batch Size & Learning Rate & Weight Decay \\
\midrule
1.7M   & $4$  & $32$  & $4$  & $128$   & $1\times10^{-2}$ & $256$  & $5\times10^{-3}$ & $1\times10^{-2}$ \\
7M     & $4$  & $128$ & $4$  & $512$   & $1\times10^{-2}$ & $512$  & $2\times10^{-3}$ & $1\times10^{-2}$ \\
20M    & $8$  & $256$ & $8$  & $1024$  & $8\times10^{-3}$ & $1024$  & $1\times10^{-3}$ & $5\times10^{-2}$ \\
64M    & $12$ & $512$ & $8$  & $2048$  & $8\times10^{-3}$ & $1024$  & $8\times10^{-4}$ & $1\times10^{-1}$ \\
125M   & $12$ & $768$ & $12$ & $2048$  & $3\times10^{-3}$ & $1024$   & $2\times10^{-4}$ & $1\times10^{-1}$ \\
210M   & $24$ & $768$ & $16$ & $2048$  & $3\times10^{-3}$ & $1024$   & $2\times10^{-4}$ & $1\times10^{-1}$ \\
\bottomrule
}
\appendix
\section{Training Details} \label{appendix_1}
We utilize separate data loaders to deal with different batch sizes between language modeling and recommendation tasks. Furthermore, the early stopping strategy is employed based on the validation loss of the recommendation task and patience of 100 steps.
We use the AdamW~\citep{loshchilov2018decoupled} with $\beta_{1} = 0.9$, $\beta_{2} = 0.98$, $\epsilon = 10^{-6}$, and Zero Redundancy Optimizer~\citep{rajbhandari2020zero}. We update the model using linear warm-up of the learning rate over the first $1\%$ steps, followed by cosine decay~\citep{loshchilov2016sgdr} to decrease the learning rate to $10\%$ of its initial value. The cosine decay is also applied to the $\lambda$ value. We leverage the automatic mixed-precision~\citep{micikevicius2018mixed} package in Pytorch~\citep{NEURIPS2019_bdbca288} to reduce training time and GPU memory usage. Gradient norm clipping~\citep{pascanu2013difficulty} is used with the max norm set to $0.1$ to stabilize training. Unless otherwise specified, all results are reported by $125\text{M}$ transformer decoder~\citep{NIPS2017_3f5ee243}. 
All models use a vocabulary size of $50,258$ and a max sequence length of $2,048$. The hyperparameter values for different sizes of \name~is presented in \cref{tab:size-param}. All the results are averaged over the 20 runs. 

\section{Details of Comparison Models} \label{appendix_2}
\noindent\textbf{Behavior Sequence Transformer} (BST)~\citep{chen2019behavior} embeds user history logs as low-dimensional vectors and passes them to the transformer layers to model underlying user preferences. 
\\\\
\noindent\textbf{LightGCN}~\citep{he2020lightgcn} leverages Graph Convolution Network~\citep{kipf2016semi} for enhancing collaborative filtering. It linearly propagates user and item embeddings of a bipartite interaction graph. The final embedding is computed by the sum of the embeddings propagated at each layer.
\\\\
\noindent\textbf{UserBERT}~\citep{10.1145/3340531.3412747} incorporates two self-supervision tasks for pretraining. These pretext tasks effectively capture the relations between user behaviors and inherent user interests. It finally finetuned models on target tasks.
\\\\
\noindent\textbf{UniSRec}~\citep{hou2022towards} proposes to combine parametric whitening and MoE adaptor for learning personalized representation. UniSRec pretrains user history by sequence-to-sequence contrastive learning and then finetunes the model to downstream tasks.
\\\\
\noindent\textbf{M6Rec}~\citep{cui2022m6} employs prompt tuning of pretrained language models for building a unified framework. M6Rec fully utilizes text inputs to generalize to any domains/systems and has the ability to perform zero-shot learning. Since they did not release pretrained M6~\citep{10.1145/3447548.3467206}, we used Huggingface RoBERTa~\citep{liu2019roberta} to implement it.$\footnote{https://huggingface.co/transformers/model\char`_doc/roberta}$
\\\\
\noindent\textbf{CLUE}~\citep{shin2021scaling} presents a plain text-based contrastive learning framework, considering heterogeneous services or applications as a modality and users as a common semantic. It then performs feature-based transfer learning for downstream tasks.

\section{Effect of Language Modeling on Local Curvature} \label{appendix_3}
\begin{figure}[t]
\begin{center}
\centerline{\includegraphics[width=1.0\columnwidth]{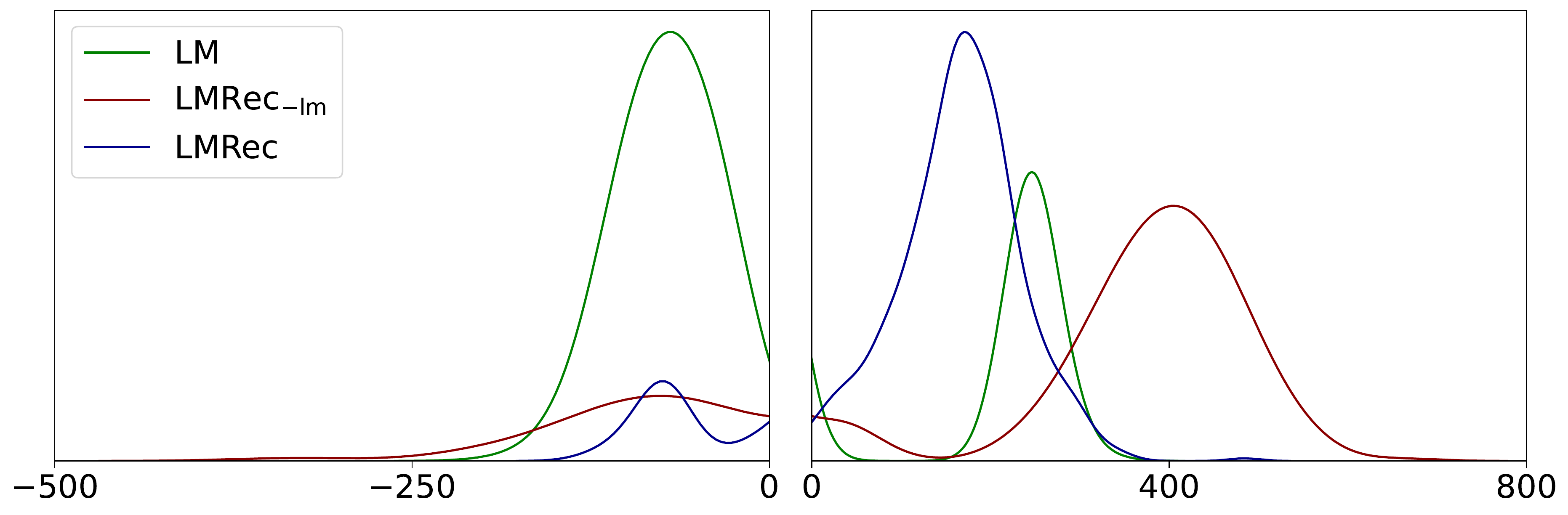}}
\caption{Hessian max eigenspectra of language model only (LM), recommendation model only (\name$_{\tiny\text{-lm}}$), and combination of them (\name) on OBS task. We calculate the Hessian max eigenvalue at the best-performing steps on downstream tasks.}
\label{fig:eigenspectra}
\end{center}
\end{figure} 
One of the most well-known criteria influencing neural network generalization is observing Hessian eigenvalues with respect to parameters. Since the Hessian is often treated as local curvature, the eigenvalues of Hessian determine the smoothness of loss landscapes. Many researchers have argued that the flat loss landscape leads to better generalization~\citep{li2018visualizing, foret2021sharpnessaware, chen2022when, park2022how}. 
We calculate and gather top-5 Hessian eigenvalues by PyHessian~\citep{yao2020pyhessian}, and resulting max eigenvalues are visualized using kernel density estimation in Scikit-learn~\citep{scikit-learn}.
Results are presented in~\cref{fig:eigenspectra}. The language model only (LM) on the OBS task produces many negative eigenvalues, which means the loss landscape is non-convex and, thus, challenging to optimize. This result is natural since the loss of the target task computed without adaptation of models cannot bring good properties. On the other hand, eigenvalues of models (\name$_{\tiny\text{-lm}}$ and \name) trained with target objectives flocked together on the positive side. The magnitude of the eigenspectra of \name model is smaller than that of \name$_{\tiny\text{-lm}}$ model. It means that learning two objectives simultaneously improves the robustness and generality of model performance on downstream tasks.

\section{Online A/B Experiment} \label{appendix_4}
We run A/B experiments on product collection recommendation tasks using \name$^{\tiny\text{TL}}_{\tiny\text{+agnostic}}$ user feature to verify the practical usage of our method online.
The product collection is a collection of products allotted by merchandisers with a particular category such as “Plush robe coats for men”, “Winter sale special offer”, and “Best backpacks for high school students”. This task is to recommend the product collection banner, linked to a page displaying a list of products. 

We pretrain~\name$^{\tiny\text{TL}}_{\tiny\text{+agnostic}}$ with OBS, OTA, and ECOMM and then transfer to the product collection recommendation (target task). The mean pooled task-specific and task-agnostic user features are used as the final user features. During the $14$ days of online experimentation, we measured two important metrics for the online recommender system, CTR and GMV, to track user satisfaction with the platform. CTR represents the click/view rate of recommendation, and GMV is the total value of sold products through recommendation. All models take the same amount of user traffic. 

\end{document}